\documentclass[12pt]{article}
\newcommand{\be}{\begin{equation}}
\newcommand{\ee}{\end{equation}}
\newcommand{\bea}{\begin{eqnarray}}
\newcommand{\eea}{\end{eqnarray}}

\newcommand{\fourth}{\raisebox{.15ex}{\scriptsize$\frac{1}{4}$}}

\newcommand{\Tr}{\mbox{\,Tr\,}}

\newcommand{\cO}{\mbox{$\cal O$}}

\newcommand{\NPB}{\mbox{Nucl. Phys. B}}
\newcommand{\PLB}{\mbox{Phys. Lett. B}}

\newcommand{\CMP}{\mbox{Comm. Math. Phys.}}

\newcommand{\PRD}{\mbox{Phys. Rev. D}}

\newcommand{\basispl}{
   \put(-.5,-.5){\line(1,0){1}}
   \put(.5,-.5){\line(0,1){1}}
   \put(.5,.5){\line(-1,0){1}}
   \put(-.5,.5){\line(0,-1){1}}}
\newcommand{\plaq}{\setlength{\unitlength}{.5cm}\raisebox{-.2cm}{
   \begin{picture}(1.2,1.2)(-.6,-.6)
   \basispl
   \put(-.5,-.5){\circle*{.2}}
   \put(-.5,.5){\circle*{.2}}
   \put(.5,-.5){\circle*{.2}}
   \put(.5,.5){\circle*{.2}}
   \put(.5,0){\vector(0,1){0}}
   \put(-.6,-.6){\makebox(0,0)[tr]{\footnotesize $x$}}
   \put(-.55,0){\makebox(0,0)[r]{\footnotesize $\nu$}}
   \put(0,-.55){\makebox(0,0)[t]{\footnotesize $\mu$}}
   \end{picture}}}
\newcommand{\twooneplaq}{\setlength{\unitlength}{.5cm}
   \raisebox{-.2cm}{
   \begin{picture}(2.2,1.2)(-1.1,-.6)
   \put(-1,-.5){\line(1,0){2}}
   \put(-1,.5){\line(1,0){2}}
   \put(-1,-.5){\line(0,1){1}}
   \put(1,-.5){\line(0,1){1}}
   \multiput(-1,-.5)(1,0){3}{\circle*{.2}}
   \multiput(-1,.5)(1,0){3}{\circle*{.2}}
   \put(-1.1,-.6){\makebox(0,0)[tr]{\footnotesize $x$}}
   \put(-1.05,0){\makebox(0,0)[r]{\footnotesize $\nu$}}
   \put(-.3,-.55){\makebox(0,0)[t]{\footnotesize $\mu$}}
   \put(1,0){\vector(0,1){0}}
   \end{picture}}}

\newcommand{\cornplaq}{\setlength{\unitlength}{.5cm}
   \raisebox{-.3268cm}{
   \begin{picture}(1.7071,1.7071)(-.7071,-.7071)
   \put(-.7071,-.7071){\line(0,1){1}}
   \put(0,1){\line(1,0){1}}
   \put(1,1){\line(0,-1){1}}
   \put(-.7071,-.7071){\line(1,0){1}}
   \put(0,1){\line(-1,-1){.7071}}
   \put(1,0){\line(-1,-1){.7071}}
   \put(-.7071,-.7071){\circle*{.1}}
   \put(-.7071,.2929){\circle*{.2}}
   \multiput(0,0)(1,0){2}{\circle*{.2}}
   \multiput(0,1)(1,0){2}{\circle*{.2}}
   \multiput(-.7071,-.7071)(1,0){2}{\circle*{.2}}
   \multiput(0,0)(.25,0){4}{\circle*{.03}}
   \multiput(0,0)(0,.25){4}{\circle*{.03}}
   \multiput(0,0)(-.1768,-.1768){4}{\circle*{.03}}
   \put(1,0.5){\vector(0,1){0}}
   \put(-.8,-.8){\makebox(0,0)[tr]{\footnotesize $x$}}
   \put(-.8,-.2){\makebox(0,0)[r]{\footnotesize $\nu$}}
   \put(-.2,-.8){\makebox(0,0)[t]{\footnotesize $\mu$}}
   \put(.9,-.4){\makebox(0,0)[t]{\footnotesize $\lambda$}}
   \end{picture}}}
\newcommand{\twoplaq}{\setlength{\unitlength}{1cm}\raisebox{-.5cm}{
   \begin{picture}(1.2,1.2)(-.6,-.6)
   \basispl
   \put(-.5,-.5){\circle*{.1}}
   \put(-.5,.5){\circle*{.1}}
   \put(.5,-.5){\circle*{.1}}
   \put(.5,.5){\circle*{.1}}
   \put(0,-.5){\circle*{.1}}
   \put(0,.5){\circle*{.1}}
   \put(.5,0){\circle*{.1}}
   \put(-.5,0){\circle*{.1}}
   \put(.5,-.2){\vector(0,1){0}}
   \put(-.55,-.55){\makebox(0,0)[tr]{\footnotesize $x$}}
   \put(-.55,-.2){\makebox(0,0)[r]{\footnotesize $\nu$}}
   \put(-.2,-.55){\makebox(0,0)[t]{\footnotesize $\mu$}}
   \end{picture}}}

\newcommand{\refeq}[1]{\mbox{eq.~(\ref{eq:#1})}}
\newcommand {\LW}{\mbox{L\"{u}scher-Weisz}}

\newcommand{\Journal}[4]{{#1} {#2} (#4) #3}

\begin{document}
\vskip-1cm
\hfill INLO-PUB-17/96
\vskip5mm
\begin{center}
{\LARGE{\bf{\underline{Square Symanzik action to one-loop order}}}}\\
\vspace{1cm}
{\large Jeroen Snippe} \\
\vspace{1cm}
Instituut-Lorentz for Theoretical Physics,\\
University of Leiden, PO Box 9506,\\
NL-2300 RA Leiden, The Netherlands.\\
\end{center}
\vspace*{5mm}{\narrower\narrower{\noindent
\underline{Abstract}:  We present the one-loop coefficients for an alternative
Symanzik improved lattice action with gauge groups \mbox{SU(2)} or
\mbox{SU(3)}.}}
\vspace{1.5cm}

Recently a new improved lattice action, called the square Symanzik action, was
introduced by adding a $2 \times 2$ Wilson loop to the \LW\ Symanzik
action~\cite{sym,lw}:
\bea
S(\{c_i\})&\equiv&\sum_x\Tr\Biggl\{c_0(g_0^2)
\sum_{\mu\neq\nu}\left\langle1-\plaq\
\right\rangle+2c_1(g_0^2)\sum_{\mu\neq\nu}\left\langle1-\twooneplaq\
\right\rangle\nonumber\\&&+
\frac{4}{3}c_2(g_0^2)\sum_{\mu\neq\nu\neq\lambda}\left\langle1-\cornplaq
\ \right\rangle+c_4(g_0^2)\sum_{\mu\neq\nu}\left\langle1-\twoplaq \
\right\rangle\Biggr\},
\eea
where the $<>$ imply averaging over the two opposite directions for each of the
links. The inclusion of the $2 \times 2$ loop allows a simple diagonalization
of the gauge field propagator, provided one takes $c_0 \, c_4 = c_1^2$ (where
$c_i \equiv c_i(g_0^2 = 0)$). This simplifies certain analytic calculations,
even if tadpole corrections~\cite{lm} are incorporated. For details we refer to
ref.~\cite{gbs}.

The aim of Symanzik improvement is to cancel leading ($\cO(a^2)$) corrections
in the lattice spacing $a$. The simplest choice for (on-shell) Symanzik
improvement at tree-level amounts to~\cite{lw}
\be
c_0 = 5/3,\ \ \ c_1 = -1/12,\ \ \ c_2 = 0, \ \ \ c_4 = 0.
\label{eq:LWtree}
\ee
For the square action one takes instead~\cite{gbs}
\be
c_0 = 16/9,\ \ \ c_1 = -1/9,\ \ \ c_2 = 0, \ \ \ c_4 = 1/144,
\label{eq:sqtree}
\ee
which satisfies $c_0 \, c_4 = c_1^2$. At tree-level many other Symanzik
improved actions can be easily constructed. This freedom has been used, e.g. in
ref.~\cite{alf}, to study the universality of improvement by comparing the
effectiveness of alternative actions.

Up to now there has been only one choice of the improvement coefficients,
\refeq{LWtree}, for which a one-loop calculation was completed~\cite{lw}. Here
we present our results of a one-loop calculation belonging to the square
Symanzik action, \refeq{sqtree}. Details of our calculation, that is based on
the methods of L\"{u}scher, Weisz and Wohlert~\cite{lw,ww}, will be presented
elsewhere. Introducing the notation $c_i(g_0^2) = c_i + c_i' \, g_0^2 +
\cO(g_0^4)$ we refer to table~\ref{tab:coeff} for the coefficients $c_i'$. At
this point we stress that $c_4'$ is a free parameter because in the expansion
of the action to $\cO(a^2)$ only the combinations $\tilde{c}_0(g_0^2) \equiv
c_0(g_0^2) - 16 c_4(g_0^2)$ and $\tilde{c}_1(g_0^2) \equiv c_1(g_0^2) +
4c_4(g_0^2)$ contain $c_4(g_0^2)$.

\begin{table}
\begin{center}
\begin{tabular}{|c|r@{.}ll||r@{.}ll|}
\hline
&\multicolumn{3}{c||}{\LW}&\multicolumn{3}{c|}{square}\\
\hline
$\tilde{c}_0'$ & 0&135160(13) & $(N=2)$ & 0&113417(11) & $(N=2)$\\
               & 0&23709(6)   & $(N=3)$ & 0&19320(4) &   $(N=3)$\\
\hline
$\tilde{c}_1'$ & -0&0139519(8) & $(N=2)$ & -0&0112766(7) & $(N=2)$\\
               & -0&025218(4)  & $(N=3)$ & -0&019799(2)  & $(N=3)$\\
\hline
$c_2'$ & -0&0029431(8) & $(N=2)$ & -0&0029005(7) & $(N=2)$\\
               & -0&004418(4)  & $(N=3)$ & -0&004351(2) &  $(N=3)$\\
\hline
$\Lambda/\Lambda_{\mbox{\scriptsize Wilson}}$ & 4&1308935(3) & $(N=2)$ &
4&0919901(2) & $(N=2)$\\
               & 5&2921038(3) & $(N=3)$ & 5&2089503(2) & $(N=3)$\\
\hline
$w(1,1)$     & \multicolumn{3}{c||}{0.366262680(2)} &
\multicolumn{3}{c|}{0.3587838551(1)}\\
\hline
$w(1,2)$     & \multicolumn{3}{c||}{0.662626785(2)} &
\multicolumn{3}{c|}{0.6542934512(1)}\\
\hline
$w(2,2)$     & \multicolumn{3}{c||}{1.098143594(2)} &
\multicolumn{3}{c|}{1.0887235337(1)}\\
\hline
\end{tabular}
\caption{One-loop improvement coefficients $c_i'$, defined by $c_i(g_0^2) = c_i
+ c_i'\, g_0^2 + \cO(g_0^4)$, for the \LW\ and square Symanzik actions. $N$ is
the number of colors, while $\tilde{c}_0'$ and $\tilde{c}_1'$ stand for $c_0' -
16 c_4'$ and $c_1' + 4 c_4'$ respectively. For completeness we include the
Lambda parameter ratios and the expectation values of a few  $a \times b$
Wilson loops, $\langle N^{-1} \mbox{Re Tr }U(a \times b)\rangle \equiv 1-
\protect\fourth g_0^2 (N - N^{-1}) w(a,b) + \cO(g_0^4)$. By convention the
tadpole parameter $u_0$ equals $\langle N^{-1} \mbox{Re Tr }U(1 \times
1)\rangle^{1/4}$.}
\label{tab:coeff}
\end{center}
\end{table}

The following checks were performed to convince ourselves of the validity of
the results in table~\ref{tab:coeff}.
\begin{itemize}
\item For the \LW\ action all results of the original calculation~\cite{lw,ww}
were reproduced, in most cases to a slightly higher accuracy. Especially the
agreement with ref.~\cite{lw} is non-trivial because we used covariant, instead
of coulomb, gauge fixing.
\item Coefficients are extracted from physical quantities computed as a
function of the lattice spacing. We checked that divergences cancel, the
one-loop beta function is reproduced, continuum limits are independent of the
action chosen, and $a^2\ln(a)$ terms do not appear for the \LW\ and square
actions---as expected from Symanzik's analysis for $\varphi^4$~\cite{sym}.
\item  The combination $\tilde{c}_1' - c_2'$ was computed both using the static
quark potential method of ref.~\cite{ww} and the twisted finite volume method
of ref.~\cite{lw}. The agreement is better than 0.003\%.
\item Using three completely different methods: (a) static quark potential; (b)
three point vertex in a twisted finite volume; (c) (for SU(2)) a background
field calculation in a periodic finite volume~\cite{gbs}, the Lambda parameters
extracted agree to at least six digits.
\end{itemize}

We conclude with testing how well the tadpole correction~\cite{lm} to the
SU(3)~tree-level square Symanzik action predicts the one-loop correction.
Since, to $\cO(a^2)$, $c_4(g_0^2)$ can be freely chosen, the relevant test is
comparing
\be
\frac{\tilde{c}_1(g_0^2)}{\tilde{c}_0(g_0^2)} = -\frac{1}{20} (1 + 0.1217 g_0^2
+ \cO(g_0^4))
\nonumber
\ee
to
\be
\frac{c_1 u_0^{-2} + 4 c_4 u_0^{-4}}{c_0 - 16 c_4 u_0^{-4}} = -\frac{1}{20} (1
+ 0.0957 g_0^2 + \cO(g_0^4)).
\nonumber
\ee
Here $u_0 = 1 - 0.3588 g_0^2/6$ was taken from table~\ref{tab:coeff}. It
follows that the tadpole prediction captures 79\% of the one-loop correction, a
result similar to the 76\% found for the \LW\ Symanzik action. (For SU(2) one
finds 80\% for both actions).

Of course one may consider the ratios $c_1(g_0^2)/c_0(g_0^2)$ and
$c_4(g_0^2)/c_0(g_0^2)$ separately. While for $c_4' = 0.003058$, satisfying
$c_4(g_0^2) c_0(g_0^2) = (c_1(g_0^2))^2$ to one-loop order, the tadpole
prediction is off by 21\% in both ratios, for $c_4' = 0.002401$ the deviations
are only 11\%.

\section*{Acknowledgments}

The author is grateful to Pierre van Baal and Margarita Garc\'{\i}a P\'erez for
encouraging discussions.


\begin{thebibliography}{9}
\bibitem{sym}K. Symanzik, \Journal{\NPB}{226}{187, 205}{1983}.
\bibitem{lw}M. L\"{u}scher and P. Weisz, \Journal{\PLB}{158}{250}{1985};
\Journal{\NPB}{266}{309}{1986}; \Journal{\CMP}{97}{59}{1985};
\Journal{}{98}{433 (E)}{1985}.
\bibitem{lm}G.P. Lepage and P.B. Mackenzie, \Journal{\PRD}{48}{2250}{1993}.
\bibitem{gbs} M. Garc\'{\i}a P\'erez, J. Snippe and P. van Baal,
hep-lat/9608036, submitted to \PLB,
and references therein.
\bibitem{alf}M. Alford, e.a., \Journal{\PLB}{361}{87}{1995}.
\bibitem{ww}P. Weisz and R. Wohlert, \Journal{\NPB}{236}{397}{1984};
\Journal{B}{247}{544 (E)}{1984}.
\end{thebibliography}
\end{document}